\documentclass[11pt]{article}
\usepackage[margin=1in]{geometry}
\newif\ifarxiv
\arxivtrue
\newcommand{\figwidth}{1\textwidth}

\usepackage[shortlabels]{enumitem}
\usepackage[numbers]{natbib}

\usepackage{graphicx}
\usepackage{authblk}

\usepackage{amsmath}
\usepackage{amsthm}
\usepackage{amsfonts}
\usepackage{amssymb}
\usepackage{bbm}

\usepackage[hidelinks]{hyperref}
\usepackage{xcolor}
\usepackage{cancel} 
\usepackage{fancyhdr}
\fancypagestyle{firstpagefooter}{
    \fancyhf{} 
    \fancyfoot[L]{\footnotesize © 2024 IEEE. Personal use of this material is permitted. Permission from IEEE must be obtained for all other uses, in any current or future media, including reprinting/republishing this material for advertising or promotional purposes, creating new collective works, for resale or redistribution to servers or lists, or reuse of any copyrighted component of this work in other works.}
}

\newtheorem{thm}{Theorem}[section]
\newtheorem{lemma}[thm]{Lemma}
\newtheorem{Def}[thm]{Definition}
\newtheorem{cor}[thm]{Corollary}
\newtheorem{prop}[thm]{Proposition}
\newtheorem{rmk}[thm]{Remark}
\newtheorem{assum}[thm]{Assumption}

\newcommand{\grad}{\mathrm{grad}}
\newcommand{\RD}{\mathbb{R}^N}
\newcommand{\RDp}{\mathbb{R}^N_+}
\newcommand{\RDpp}{\mathbb{R}^N_{++}}
\newcommand{\simplex}{\Delta^N_{+}}
\newcommand{\simplexpp}{\Delta^N_{++}}
\newcommand{\du}{\mathrm{d}}

\DeclareMathOperator*{\argmin}{\arg\min}

\title{Computing Augustin Information via \\  
Hybrid Geodesically Convex Optimization}
\author[1]{\normalsize Guan-Ren Wang}
\author[2]{\normalsize Chung-En Tsai}
\author[3, 4, 5, 6]{\normalsize Hao-Chung Cheng}
\author[1, 2, 4, 6]{\normalsize Yen-Huan Li}
\affil[1]{\small Graduate Institute of Networking and Multimedia, National Taiwan University}
\affil[2]{\small Department of Computer Science and Information Engineering, National Taiwan University}
\affil[3]{\small Department of Electrical Engineering and Graduate Institute of Communication Engineering, National Taiwan University}
\affil[4]{\small Department of Mathematics, National Taiwan University}
\affil[5]{\small Hon Hai (Foxconn) Quantum Computing Centre}
\affil[6]{\small Center for Quantum Science and Engineering, National Taiwan University}

\date{}

\begin{document}

\maketitle

\thispagestyle{firstpagefooter}
\begin{abstract}
   We propose a Riemannian gradient descent with the Poincaré metric to compute the order-$\alpha$ Augustin information, a widely used quantity for characterizing exponential error behaviors in information theory.
   We prove that the algorithm converges to the optimum at a rate of $\mathcal{O}(1 / T)$.
   As far as we know, this is the first algorithm with a non-asymptotic optimization error guarantee for all positive orders.
   Numerical experimental results demonstrate the empirical efficiency of the algorithm.
   Our result is based on a novel hybrid analysis of Riemannian gradient descent for functions that are geodesically convex in a Riemannian metric and geodesically smooth in another.
\end{abstract}
\section{Introduction}\label{sec:intro}

Characterizing operational quantities of interest in terms of a single-letter information-theoretic quantity is fundamental to information theory.
However, certain single-letter quantities involve an optimization that is not easy to solve.
This work aims to propose an optimization framework for computing an essential family of information-theoretic quantities, termed the \emph{Augustin information} \cite{augustin69, augustin1978noisy, Csi95, Ver15, nakibouglu2019augustin, CGH18, Ver21, CN21}, with the first non-asymptotic optimization error guarantee.

Consider two probability distributions $P$ and $Q$ sharing the same finite alphabet $\mathcal{Y}$.
The order-$\alpha$ Rényi divergence ($\alpha \in (0,\infty) \backslash \{1\}$) is defined as \cite{Ren62, EH14}
\begin{align}
\notag
D_\alpha(P\Vert Q):=\frac{1}{\alpha-1}\log\sum_{y\in\mathcal{Y}}P^\alpha(y)Q^{1-\alpha}(y).
\end{align}
As $\alpha$ tends to $1$, we have
\begin{align*}
    &\lim_{\alpha\to 1}D_\alpha(P\Vert Q)=D(P\Vert Q),
\end{align*}
where $D(P\Vert Q) := \mathbb{E}_{Y\sim P}\left[ \log \frac{P(Y)}{Q(Y)} \right]$ is the Kullback--Leibler (KL) divergence.
The KL divergence defines Shannon's mutual information for a prior distribution $P_X$ on the input alphabet $\mathcal{X}$ and a conditional distribution $P_{Y|X}$ as
\begin{align}
\notag
I( P_X, P_{Y|X})
&:= \min_{Q_Y}\mathbb{E}_X D(P_{Y\mid X}(\cdot\mid X)\Vert Q_Y)\\
\notag
&= D(P_{Y\mid X}\Vert P_Y | P_X),
\end{align}
where the minimization is over all probability distributions on $\mathcal{Y}$ and $P_Y = \mathbb{E}_X[ P_{Y|X}]$.
Augustin generalized Shannon's mutual information to the order-$\alpha$ Augustin information \cite{augustin69, augustin1978noisy}, defined as
\begin{equation}\label{eq:Augustin information}
    I_{\alpha}( P_X, P_{Y|X}):=\min_{Q_Y}\mathbb{E}_X D_{\alpha}(P_{Y\mid X}(\cdot\mid X)\Vert Q_Y).
\end{equation}

The Augustin information yields a wealth of applications in information theory, including:
\begin{enumerate}[(i)]
    \item characterizing the cut-off rate of channel coding \cite{Csi95},
    
    \item determining the error exponent of optimal constant compositions codes for rates below Shannon's capacity \cite{Har68, augustin69, Bla74, augustin1978noisy, CK11, nakibouglu2020simple, DW14, CHT19},
    
    \item determining the strong converse exponent of channel coding for rates above Shannon's capacity
    \cite{arimoto73, omura75, dueckK79, mosonyiO17, mosonyiO21},

    \item determining the error exponent of variable-length data compression with side information 
    \cite{chen_reliability_2017, CHDH22},

    \item demonstrating an achievable secrecy exponent for privacy amplification and wiretap channel coding via regular random binning \cite{mojahedian2019correlation, SGC23}.
\end{enumerate}

In spite of the operational significance of the Augustin information, computing it is not an easy task.
The order-$\alpha$ Augustin information does not have a closed-form expression in general unless $\alpha = 1$.
The optimization problem \eqref{eq:Augustin information} is indeed convex \cite[Theorem 12]{EH14}, so we may consider solving it via gradient descent.
However, existing non-asymptotic analyses of gradient descent assume either Lipschitzness or smoothness\footnote{The term ``smooth'' used in this paper does not mean infinite differentiability. We will define smoothness in Section \ref{def:smoothness}.} of the objective function
\cite{bubeck2015convex, nesterov2018lectures, lan2020first}.
These two conditions, which we refer to as smoothness-type conditions, are violated by the optimization problem \eqref{eq:Augustin information} \cite{you2022minimizing}.

In this paper, we adopt a \emph{geodesically convex optimization} approach \cite{zhang2016first, boumal2023introduction, vishnoi2018geodesic, absil2008optimization}.
This approach utilizes generalizations of convexity and smoothness-type conditions for Riemannian manifolds, namely geodesic convexity, geodesic Lipschitzness, and geodesic smoothness \cite{zhang2016first, rapcsak1991geodesic, udriste1994convex, wiesel2012geodesic, zhang2013multivariate, hosseini2015matrix, hosseini2020alternative, chewi2020gradient, antonakopoulos2020online}.
Since the lack of standard smoothness-type conditions is the main difficulty in applying gradient descent-type methods, we may find an appropriate Riemannian metric with respect to which the objective function in the optimization problem \eqref{eq:Augustin information} is geodesically smooth.

Indeed, we have found that the objective function is geodesically smooth with respect to a Riemannian metric called the Poincaré metric.
However, numerical experiments suggest that the objective function is geodesically \emph{non-convex} under this metric.
This poses another challenge.
Existing analyses for Riemannian gradient descent, a direct extension of vanilla gradient descent, require both geodesic convexity and geodesic smoothness under the same Riemannian metric \cite{zhang2016first, boumal2023introduction}.

Our main theoretical contribution lies in analyzing Riemannian gradient descent under a \emph{hybrid} scenario, where the objective function is geodesically convex with respect to one Riemannian metric and geodesically smooth with respect to another.
We prove that under this hybrid scenario, Riemannian gradient descent converges at a rate of $\mathcal{O}(1/T)$, identical to the rate of vanilla gradient descent under the Euclidean setup.

Given that the objective function for computing the Augustin information is convex under the standard Euclidean structure and geodesically smooth under the Poincaré metric, our proposed hybrid framework offers a viable approach to solve this problem.
In particular, we prove that Riemannian gradient descent with the Poincaré metric converges at a rate of $\mathcal{O}(1/T)$ for computing the Augustin information.
This marks the first algorithm for computing the Augustin information that has a non-asymptotic optimization error bound for all $\alpha>0$.

\noindent\textbf{Notations.}\quad We denote the all-ones vector by $\mathbbm{1}$.
We denote the set of vectors with non-negative entries and strictly positive entries by $\RDp$ and $\RDpp$, respectively.
The $i$-th entry of $x$ is denoted by $x^{(i)}$.
For any function $f:\mathbb{R}\to\mathbb{R}$ and vector $x$, we define $f(x)$ as the vector whose $i$-th entry equals $f(x^{(i)})$.
Hence, for example, the $i$-th entry of $\exp(x)$ is $\exp x^{(i)}$. 
The probability simplex in $\RDp$ is denoted by $\simplex$, and the intersection of $\RDpp$ and $\simplex$ is denoted by $\simplexpp$.
The notations $\odot$ and $\oslash$ denote entry-wise product and entry-wise division, respectively.
For $x, y\in\RD$, the partial order $x\leq y$ means $y-x\in\RDp$.
For a set $S\in\RD$, the boundary of $S$ is denoted by $\partial S$.
The set $\{1,2,\ldots, N\}$ is denoted by $[N]$.

\section{Related Work}\label{sec:realted work}

\subsection{Computing the Augustin Information}
\citet{nakibouglu2019augustin} showed that the minimizer of the optimization problem \eqref{eq:Augustin information} satisfies a fixed-point equation and proved that the associated fixed-point iteration converges to the order-$\alpha$ Augustin information for $\alpha \in (0, 1)$.
\citet[Lemma 6]{CN21} provided a descent lemma for a specific update rule, which implies asymptotic convergence for the optimization problem \eqref{eq:Augustin information} for all $\alpha >0$.
\citet{li2019convergence} provided a line search gradient-based algorithm to solve optimization problems on the probability simplex, which can also be used to solve the optimization problem \eqref{eq:Augustin information} for all $\alpha  > 0$.
\citet{you2022minimizing} proposed a gradient-based method for minimizing quantum Rényi divergences, which includes the optimization problem \eqref{eq:Augustin information} as a special case for all $\alpha>0$.
These works only guarantee asymptotic convergence of the algorithms.
Analysis of the convergence rate before the present work is still missing.

\subsection{Hybrid Analysis}
\citet{antonakopoulos2020online} considered (Euclidean) convex and geodesically Lipschitz objective functions, analyzing the regret rates of the follow-the-regularized-leader algorithm and online mirror descent.
\citet{weber2023global} considered geodesically convex and (Euclidean) smooth objective functions, analyzing the optimization error of the convex-concave procedure.
The former did not consider the hybrid case of geodesic convexity and geodesic smoothness under different Riemannian metrics.
The latter only analyzed a special case of the hybrid scenario we proposed.
Neither of them analyzed Riemannian gradient descent.
\section{Preliminaries}\label{sec:preliminary}
This chapter introduces relevant concepts in geodesic convex optimization.

\subsection{Smooth Convex Optimization}\label{sec:convex optimization}
If an optimization problem is convex, we may consider using (projected) gradient descent to approximate a global minimizer.
Existing non-asymptotic error bounds of gradient descent require the objective function to be either Lipschitz or smooth.
\begin{Def}\label{def:smoothness}
    We say a function $f:\RD\to\mathbb{R}$ is $L$-Lipschitz for some $L>0$ if for any $x, y\in\RD$, the function satisfies
    \begin{equation*}
        |f(y)-f(x)|\leq L\Vert y-x\Vert_2.
    \end{equation*}
    We say a function $f:\RD\to\mathbb{R}$ is $L$-smooth for some $L>0$ if for any $x, y\in\RD$, the function satisfies
    \begin{equation*}
        f(y)\leq f(x)+\langle\nabla f(x), y-x\rangle+\frac{L}{2}\Vert y-x\Vert_2^2.
    \end{equation*}
\end{Def}

Given a convex optimization problem, gradient descent converges at a rate of $\mathcal{O}(1/\sqrt{T})$ for the Lipschitz case and $\mathcal{O}(1/T)$ for the smoothness case, where $T$ denotes the number of iterations.
However, \citet[Proposition III.1]{you2022minimizing} showed that the objective function in the optimization problem \eqref{eq:Augustin information} is neither Lipschitz nor smooth.

\subsection{Basics of Riemannian Geometry}\label{sec:Riemannian geometry}
An $N$-dimensional topological manifold $\mathcal{M}$ is a topological space that is Hausdorff, second countable, and locally homeomorphic to $\RD$.
The tangent space at a point $x \in \mathcal{M}$, denoted as $T_x\mathcal{M}$, is the set of all vectors tangent to the manifold at $x$.
A Riemannian metric $\mathfrak{g}$ on $\mathcal{M}$ defines an inner product on each tangent space.
The inner product on $T_x\mathcal{M}$ is denoted as $\langle\cdot, \cdot\rangle_x$.
A Riemannian manifold $(\mathcal{M}, \mathfrak{g})$ is a topological manifold $\mathcal{M}$ equipped with a Riemannian metric $\mathfrak{g}$.
Given a Riemannian metric, the induced norm of a tangent vector $v\in T_x\mathcal{M}$ is given by $\Vert v\Vert_x:=\sqrt{\langle v, v\rangle_x}$.
The length of a curve $\gamma:[0, 1]\to\mathcal{M}$ is defined as $L(\gamma):=\int_0^1\Vert \gamma'(t)\Vert_{\gamma(t)}\mathrm{d}t$, and the induced distance between two points $x, y\in\mathcal{M}$ is given by $d(x, y)=\inf_{\gamma}L(\gamma)$, where the infimum is taken over all curves $\gamma$ connecting $x$ and $y$.
A geodesic is a curve connecting $x$ and $y$ with constant speed and satisfying $L(\gamma)=d(x, y)$.
For each $x\in\mathcal{M}$ and $v\in T_x\mathcal{M}$, there is a unique geodesic $\gamma_v:[0, 1]\to\mathcal{M}$ such that $\gamma_v(0)=x$ and $\gamma_v'(0)=v$ \cite[Corollary 4.28]{lee2018introduction}.
The exponential map at a point $x\in\mathcal{M}$ is the map $\exp_x:T_x\mathcal{M}\to\mathcal{M}$ such that $\exp_x(v)=\gamma_v(1)$.\footnote{In general, the exponential map may only be locally defined, while the exponential map considered in this paper is defined on the whole tangent space.}
The logarithmic map $\log_x:\mathcal{M}\to T_x\mathcal{M}$ at a point $x\in\mathcal{M}$ is the inverse of $\exp_x$.
Given a differentiable function $f:\mathcal{M}\to\mathbb{R}$, the Riemannian gradient of $f$ at $x\in\mathcal{M}$ is the unique tangent vector $\grad f(x)$ that satisfies $\langle \grad f(x), v\rangle_x=\du f(x)[v]$ for all $v\in T_x\mathcal{M}$.

\subsection{Geodesically Convex Optimization}\label{sec:geodesic convex optimization}
The notions of convexity and smoothness can be extended for functions defined on Riemannian manifolds.

\begin{Def}[Geodesic convexity, geodesic smoothness \cite{zhang2016first}]
    Let $f:\mathcal{M}\to\mathbb{R}$ be a function defined on a Riemannian manifold $(\mathcal{M}, \mathfrak{g})$.
    We say the function $f$ is geodesically convex (g-convex) on $(\mathcal{M}, \mathfrak{g})$ if for every $x, y\in\mathcal{M}$, the function satisfies
    \begin{equation*}
        f(y)\geq f(x)+\langle \grad f(x), \log_x y\rangle_x,
    \end{equation*}
    We say the function $f$ is geodesically $L$-smooth (g-$L$-smooth) on $(\mathcal{M}, \mathfrak{g})$ if for every $x, y\in\mathcal{M}$, the function satisfies
    \begin{equation*}
        f(y)\leq f(x)+\langle \grad f(x), \log_x y \rangle_x+\frac{L}{2}d^2(x, y),
    \end{equation*}
    for some $L\geq 0$, where $d$ is the induced distance.
\end{Def}

With the notions of the exponential map and Riemannian gradient, vanilla gradient descent can be extended to the so-called Riemannian gradient descent (RGD) \cite{zhang2016first, boumal2023introduction, chewi2020gradient}, which iterates as follows
\begin{equation}\label{eq:RGD}
    x_{t+1}\leftarrow \exp_{x_t}(-\eta \grad f(x_t)),
\end{equation}
where $\eta$ denotes the step size.

For Riemannian gradient descent, there exists a non-asymptotic optimization error bound analogous to that in gradient descent
\cite{zhang2016first}.

\begin{thm}[{\cite[Theorem 13]{zhang2016first}}]
    If the function to be minimized is g-convex and g-$L$-smooth on a Riemannian manifold $(\mathcal{M}, \mathfrak{g})$, then RGD with step size $\eta=1/L$ converges at a rate of $\mathcal{O}(1/T)$.
\end{thm}
\section{Hybrid Analysis of Geodesically Smooth Convex Optimization}\label{sec:main}

Consider the optimization problem
\begin{equation*}
    \min_{x\in\mathcal{M}}f(x),
\end{equation*}
where $\mathcal{M}$ is a Riemannian manifold equipped with Riemannian metrics $\mathfrak{g}$ and $\mathfrak{h}$, and $f\in\mathcal{C}^1(\mathcal{M})$ is a function lower bounded on $\mathcal{M}$.
In this paper, we consider a hybrid geodesic optimization framework in which $f$ is g-convex and g-smooth under different Riemannian metrics.

\begin{assum}\label{assum:hybrid}
    The function $f$ is g-convex with respect to the Riemannian metric $\mathfrak{g}$ and g-$L$-smooth with respect to the Riemannian metric $\mathfrak{h}$, i.e.,
    \begin{align*}
        f(y)&\geq f(x)+\langle \grad_{\mathfrak{g}} f(x), \log_{\mathfrak{g}(x)} y\rangle_{{\mathfrak{g}}(x)}\\
        f(y)&\leq f(x)+\langle \grad_{\mathfrak{h}} f(x), \log_{\mathfrak{h}(x)} y\rangle_{{\mathfrak{h}}(x)}+\frac{L}{2}d^2_{\mathfrak{h}}(x, y),
    \end{align*}
    where the subscripts $\mathfrak{g}$ and $\mathfrak{h}$ in the $\grad$, $\log$, $\langle\cdot,\cdot\rangle$, and $d(\cdot, \cdot)$ respectively indicate the specific metric associated with each geometric quantity.
\end{assum}

In this section, we prove that RGD with $\mathfrak{h}$ still converges at a rate of $\mathcal{O}(1/T)$ under Assumption \ref{assum:hybrid}.

\begin{lemma}[{\cite[Corollary 4.8]{boumal2023introduction}}] \label{lemma:descent}
    Let $f$ be g-$L$-smooth with respect to $\mathfrak{h}$ and let $x_+=\exp_{\mathfrak{h}(x)}\left(-\frac{1}{L}\grad_\mathfrak{h} f(x)\right)$ for some $x\in\mathcal{M}$.
    Then,
    \begin{equation*}
        f(x_+)-f(x) \leq -\frac{1}{2L}\Vert \grad_\mathfrak{h} f(x)\Vert_{\mathfrak{h}(x)}^2,
    \end{equation*}
    where $\exp_{\mathfrak{h}(x)}(\cdot)$, $\grad_\mathfrak{h} f(\cdot)$, and $\Vert \cdot\Vert_{\mathfrak{h}(x)}$ are induced by $\mathfrak{h}$.
\end{lemma}

With Lemma \ref{lemma:descent}, we can expect that RGD generates a convergent sequence for which the Riemannian gradient at the limit point vanishes.

\begin{lemma}[{\cite[Corollary 4.9]{boumal2023introduction}}]\label{lemma:vanishing grad}
    Let $f$ be g-$L$-smooth with respect to $\mathfrak{h}$.
    Let $\{x_t\}$ be the iterates generated by RGD with the Riemannian metric $\mathfrak{h}$, initial iterate $x_1\in\mathcal{M}$ and step size $\eta=1/L$.
    Then, we have
    \begin{equation*}
        \lim_{t\to\infty}\Vert\grad_{\mathfrak{h}} f(x_t)\Vert_{\mathfrak{h}(x_t)} = 0.
    \end{equation*}
\end{lemma}

The distance between two consecutive RGD iterates, $x$ and $x_+=\exp_{\mathfrak{h}(x)}\left(-\frac{1}{L}\grad_\mathfrak{h} f(x)\right)$, goes to zero.
This is due to the fact that $d_\mathfrak{h}(x, y)=\Vert \log_{\mathfrak{h}(x)} y\Vert_{\mathfrak{h}(x)}$ for any $x, y\in\mathcal{M}$, and the following equation:
\begin{equation*}
    \log_{\mathfrak{h}(x)}x_+=-\frac{1}{L}\grad_\mathfrak{h}f(x).
\end{equation*}
Since the Riemannian gradients at the iterates vanish, as stated in Lemma \ref{lemma:vanishing grad}, the following corollary applies.

\begin{cor}
    The limit point $x_\infty:=\lim_{t\to\infty}x_t$ of the RGD iterates in Lemma \ref{lemma:vanishing grad} exists.
\end{cor}

Then, if $x_\infty\in\mathcal{M}$, the Riemannian gradient (with respect to the Riemannian metric $\mathfrak{h}$) at $x_\infty$ is zero.
This means the iterates converge to a minimizer of $f$.

\begin{lemma}\label{lemma:asymptotic convergence}
    Suppose that Assumption \ref{assum:hybrid} holds.
    Let $\{x_t\}$ be the iterates generated by RGD with the Riemannian metric $\mathfrak{h}$, initial iterate $x_1\in\mathcal{M}$ and step size $\eta=1/L$.
    If the limit point $x_\infty\in\mathcal{M}$, then $x_\infty$ is a minimizer of $f$ on $\mathcal{M}$.
\end{lemma}

The following theorem presents the main result of this section, a non-asymptotic error bound for RGD under the hybrid geodesic optimization framework.

\begin{thm}\label{thm:main}
    Suppose that Assumption \ref{assum:hybrid} holds.
    Let $\{x_t\}$ be the iterates generated by RGD with the Riemannian metric $\mathfrak{h}$, initial iterate $x_1\in\mathcal{M}$ and step size $\eta=1/L$.
    For any $T\in\mathbb{N}$, we have
\begin{equation*}
    f(x_{T+1})-f(x_\infty)\leq\frac{2L}{T}\sup_{t\in\mathbb{N}}\Vert \log_{\mathfrak{g}(x_t)}x_\infty\Vert^2_{\mathfrak{h}(x_t)},
\end{equation*}
where $x_\infty:=\lim_{t\to\infty}x_t$.
\end{thm}

We sketch the proof below.
\begin{IEEEproof}
    Let $\delta_t:=f(x_t)-f(x_\infty)$.
    By Lemma \ref{lemma:descent}, we have
    \begin{equation*}
        \delta_{t+1}-\delta_t\leq -\frac{1}{2L}\Vert \grad_{\mathfrak{h}} f(x_t)\Vert_{\mathfrak{h}(x_t)}^2.
    \end{equation*}
    By the g-convexity of $f$ and the Cauchy-Schwarz inequalities, we write
    \begin{align*}
        \delta_t&\leq \langle -\grad_{\mathfrak{g}} f(x_t), \log_{\mathfrak{g}(x_t)}x_\infty\rangle_{\mathfrak{g}(x_t)}\\
        &=-\du f(x_t)[\log_{\mathfrak{g}(x_t)}x_\infty]\\
        &=\langle -\grad_{\mathfrak{h}} f(x_t), \log_{\mathfrak{g}(x_t)}x_\infty\rangle_{\mathfrak{h}(x_t)}\\
        &\leq \Vert\grad_{\mathfrak{h}} f(x_t)\Vert_{\mathfrak{h}(x_t)}\Vert \log_{\mathfrak{g}(x_t)}x_\infty\Vert_{\mathfrak{h}(x_t)}.
    \end{align*}
    Combining the two inequalities above, we obtain
    \begin{equation*}
        \delta_{t+1}-\delta_t\leq-\frac{1}{2L}\frac{\delta_t^2}{\Vert \log_{\mathfrak{g}(x_t)}x_\infty\Vert^2_{\mathfrak{h}(x_t)}}.
    \end{equation*}
    Then, we can follow the standard analysis of gradient descent \cite[Section 3.2]{bubeck2015convex}.
    \ifarxiv
        The rest of the proof can be found in Appendix \ref{appendix:main_proof}.
    \fi
\end{IEEEproof}

\begin{rmk}\label{rmk:rmk}
    Note that $\sup_{t\in\mathbb{N}}\Vert \log_{\mathfrak{g}(x_t)}x_\infty \Vert_{\mathfrak{h}(x_t)}$ is bounded since $\lim_{t\to\infty}x_t=x_\infty$.
\end{rmk}
\section{Application: Computing Augustin Information}\label{sec:application}

In this section, we apply Theorem \ref{thm:main} to compute the order-$\alpha$ Augustin information \eqref{eq:Augustin information}.
We begin by introducing a Riemannian metric called the Poincaré metric.
We then show that the objective function of the optimization problem \eqref{eq:Augustin information} is g-$|1-\alpha|$-smooth with respect to this Riemannian metric.
Finally, we apply the main result, Theorem \ref{thm:main}, with a minor adjustment regarding a boundary issue.

\subsection{Poincaré Metric}
Consider the manifold $\mathcal{M}=\RDpp$.
Let $x\in\mathcal{M}$ and $u, v\in T_x\mathcal{M}$.
The Poincaré metric\footnote{This metric is not the exact Poincaré metric in standard differential geometry literature (see e.g., \cite{lee2018introduction}) but rather a variation of it.
We use this term for convenience, as \citet{antonakopoulos2020online} did.} is given by
\begin{equation*}
    \langle u, v\rangle_x = \langle u\oslash x, v\oslash x \rangle.
\end{equation*}

\begin{prop}[\hspace{1sp}{\cite{bhatia2019bures, pennec2006riemannian, pennec2020manifold}}]
    Given $x, y\in\RDpp$ and $v\in\RD$, we have the following:
    \begin{itemize}
        \item Riemannian distance: $d^2(x, y)=\Vert \log(y\oslash x)\Vert ^2_2$.
        \item Geodesic: $\gamma(t)=x^{1-t}\odot y^t$, which connects $x$ and $y$.
        \item Exponential map: $\exp_x(v)=x\odot\exp(v\oslash x)$.
        \item Riemannian gradient: $\grad f(x)=x^2\odot\nabla f(x)$.
    \end{itemize}
\end{prop}

\subsection{Geodesic Smoothness of the Objective Function}
Since we consider the finite alphabet $\mathcal{Y}$ case, we can identify $P_{Y|X}$ and $Q_Y$ as vectors in $\simplex$.
Denote $Q_Y$ by $x$ and view $P_{Y|X}$ as a random variable $p$ in $\simplex$.
Then, the order-$\alpha$ Augustin information \eqref{eq:Augustin information} for $p$ can be written as
\begin{equation}\label{eq:AI rewritten}
    \min_{x\in\simplex}f_{\alpha}(x),\quad f_{\alpha}(x):=\mathbb{E}_{p}D_{\alpha}(p\Vert x),
\end{equation}
where $\alpha \in \mathbb{R}_+\setminus\{1\}$ and $D_{\alpha}(y\Vert x):=\frac{1}{\alpha-1}\log\langle y^{\alpha}, x^{1-\alpha}\rangle$ is the order-$\alpha$ Rényi divergence between two probability vectors $x, y \in \simplex$.

The constraint set in the optimization problem \eqref{eq:AI rewritten} is the probability simplex $\simplex$, while the Poincaré metric is defined over the whole positive orthant $\RDpp$.
We address this inconsistency by the following lemma.
Define
\begin{equation*}
    g_{\alpha}(x):=\langle \mathbbm{1}, x\rangle+f_{\alpha}(x).
\end{equation*}

\begin{lemma}\label{lemma:relaxation}
    Let $f_\alpha$ and $g_\alpha$ be defined as above, we have
    $$\argmin_{x\in\RDp}g_{\alpha}(x)=\argmin_{y\in\simplex}f_{\alpha}(y).$$
\end{lemma}

\begin{lemma}\label{lemma:AI g-smoothness}
    The function $f_{\alpha}(x)$ is geodesically $|1-\alpha|$-smooth in $\RDpp$ with respect to the Poincaré metric; the function $g_{\alpha}(x)$ is geodesically $(|1-\alpha|+1)$-smooth on the set $\{x\in\RDpp: x\leq \mathbbm{1}\}$ with respect to the Poincaré metric.
\end{lemma}

With Lemma \ref{lemma:AI g-smoothness}, we may consider minimizing $g_\alpha$ on $\RDp$ via RGD with the Poincaré metric and step size $\eta = 1/(|1-\alpha|+1)$. Starting with $x_1 \in \RDp$, the iteration rule proceeds as
\begin{equation}\label{eq:RGD for g}
        x_{t+1}=x_t\odot\exp\left(-\frac{1}{|1-\alpha|+1}x_t\odot\nabla g_\alpha(x_t)\right).
\end{equation}

Note that we additionally require that $x\leq\mathbbm{1}$ for the g-smoothness of $g_\alpha$.
We justify this requirement by the following lemma.

\begin{lemma}\label{lemma:remain in box}
    The iterates $\{x_t\}$ generated by the iteration rule \eqref{eq:RGD for g} satisfies $x_t \leq \mathbbm{1}$.
\end{lemma}

Since the function $\langle \mathbbm{1}, x\rangle$ is convex in the Euclidean sense, the objective function $g_{\alpha}$ is also convex in the Euclidean sense.
It is desirable to apply Theorem \ref{thm:main} to $g_{\alpha}$ with $\mathfrak{g}$ as the standard Euclidean metric and $\mathfrak{h}$ as the Poincaré metric.
However, the Poincaré metric is only defined on the interior of the constraint set $\RDp$, whereas RGD for the function $g_\alpha(x)$ may yield iterates that violate the assumption that $\lim_{t\to\infty}x_t\in\mathcal{M}$ of Lemma \ref{lemma:asymptotic convergence}.
We show that even if the limit point falls on the boundary of $\RDpp$, the limit point is still a minimizer when considering the Poincaré metric on $\RDpp$.

\begin{lemma}\label{lemma:boundary issue}
    Let $f\in\mathcal{C}^1(\RDp)$ be Euclidean convex and g-$L$-smooth with respect to the Poincaré metric.
    Let $\{x_t\}$ be the iterates generated by RGD with the Poincaré metric, initial iterate $x_1\in\RDpp$ and step size $\eta=1/L$.
    Then, the limit point $x_\infty:=\lim_{t\to\infty}x_t\in\RDp$ is a minimizer of $f$ on $\RDp$.
\end{lemma}

Consequently, we still have a non-asymptotic error bound for the optimization problem $\min_{x\in\RDp}g_{\alpha}(x)$.

Since the iteration rule \eqref{eq:RGD for g} is for the optimization problem $\min_{x\in\RDp}g_\alpha(x)$, these iterates may fall outside the constraint set $\simplex$ of the original problem.
We show that the sequence of normalized iterates still converges at a rate of  $\mathcal{O}(1/T)$.

\begin{prop}\label{prop:applied to f}
    Let $\{x_t\}$ be the iterates generated by the iteration rule \eqref{eq:RGD for g} with $x_1\in\simplex$.
    For any $T\in\mathbb{N}$, we have
    \begin{equation*}
        f_\alpha(\overline{x}_{T+1})-f_\alpha(x^\star)\leq\frac{2(|1-\alpha|+1)}{T}\sup_{t\in\mathbb{N}}\Vert x^\star\oslash x_t-\mathbbm{1}\Vert_2^2,
    \end{equation*}
    where $x^\star\in\argmin_{x\in\simplex}f_\alpha(x)$ and $\bar{x}:=\frac{x}{\Vert x \Vert_1}$ for any vector $x\in\RDp$.
\end{prop}

Note that the term $\sup_{t\in\mathbb{N}}\Vert x^\star\oslash x_t - \mathbbm{1} \Vert^2_2$ is bounded since $\lim_{t\to\infty}x_t=x^\star$ (Remark \ref{rmk:rmk}).

\subsection{Numerical Results}
We conducted numerical experiments\footnote{The code is available on GitHub at the following repository: https://github.com/CMGRWang/Computing-Augustin-Information-via-Hybrid-Geodesically-Convex-Optimization.} on the optimization problem \eqref{eq:Augustin information} by RGD, using the explicit iteration rule \eqref{eq:RGD for g}.
The experimental setting is as follows: the support cardinality of $p$ is $2^{14}$, and the parameter dimension $N$ is $2^{4}$.
The random variable $p$ is uniformly generated from the probability simplex $\simplex$.
We implemented all methods in Python on a machine equipped with an Intel(R) Core(TM) i7-9700 CPU running at 3.00GHz and 16.0GB memory.
The elapsed time represents the actual running time of each algorithm.

\begin{figure}[ht]
    \centering
    \includegraphics[width=\figwidth]{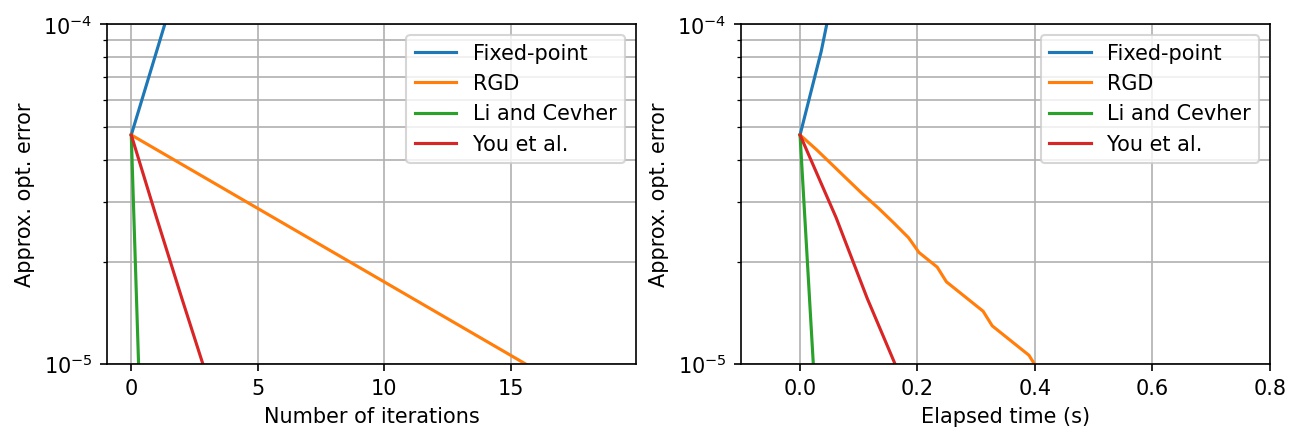}
    \caption{Convergence speeds for computing the order-$3$ Augustin information.}
    \label{fig:numerical}
\end{figure}
\begin{figure}[ht]
    \centering
    \includegraphics[width=\figwidth]{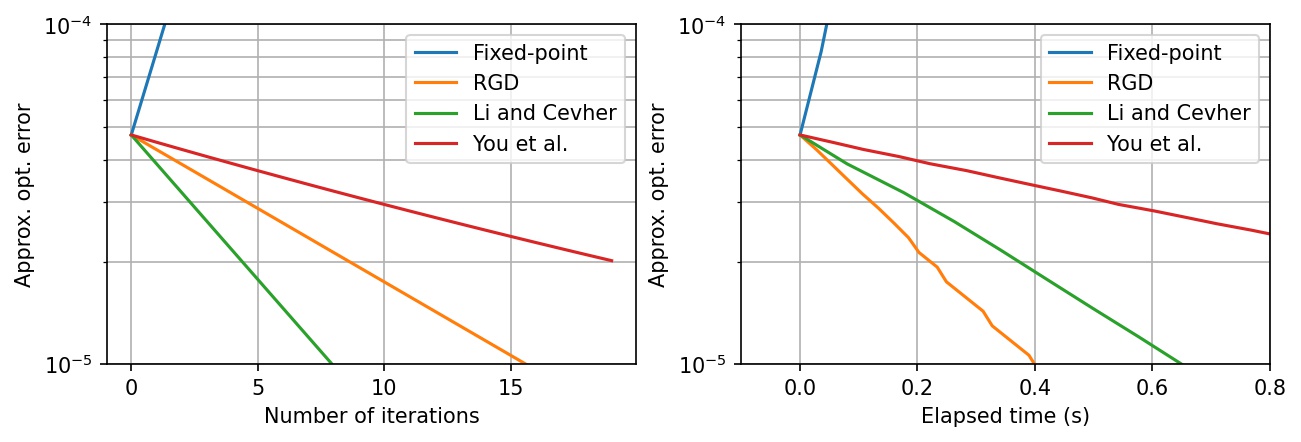}
    \caption{Slower convergence speeds for the methods proposed by \citet{li2019convergence} and by \citet{you2022minimizing}.}
    \label{fig:bad}
\end{figure}


Figure \ref{fig:numerical} demonstrates the results of computing the order-$3$ Augustin information using RGD, the fixed-point iteration in Nakıboğlu's paper \cite{nakibouglu2019augustin}, the method proposed by \citet{li2019convergence}, and the method proposed by \citet{you2022minimizing}.
We tuned the parameters for the latter two methods to achieve faster convergence speeds.
We set $\bar{\alpha}=0.4$, $r=0.7$, $\tau=0.5$ for the method proposed by \citet{li2019convergence}, and set $\delta_1=1$, $\delta=10^{-5}$, $\beta=0.99$, $\gamma=1.25$, $c=10$ for the method proposed by \citet{you2022minimizing}.
We approximate the optimal value based on the result obtained from the method proposed by \citet{li2019convergence} after $20$ iterations.

The numerical results validate our theoretical analysis, showing that RGD converges to the optimum.
The fixed-point iteration proposed by \citet{nakibouglu2019augustin} diverges, as it is only guaranteed to converge for $\alpha<1$.
We observe that RGD is slower than the methods proposed by \citet{li2019convergence} and by \citet{you2022minimizing}.
However, the latter two methods guarantee only asymptotic convergence, and tuning parameters is required to achieve faster convergence.
We found that slightly adjusting parameters can result in slow convergence speed for these two algorithms.
Figure \ref{fig:bad} shows that if we change $\bar{\alpha}$ from $0.4$ to $0.04$ and $c$ from $10$ to $100$, both methods become slower than RGD in terms of elapsed time.
Directly comparing numerical results would be unfair, given that these two methods lack convergence rate guarantees and require tuning to get better results.
\section{Conclusion}\label{sec:conclusion}
We have presented an algorithm for computing the order-$\alpha$ Augustin information with a non-asymptotic optimization error guarantee.
We have shown that the objective function of the corresponding optimization problem is geodesically $|1-\alpha|$-smooth with respect to the Poincaré metric.
With this observation, we propose a hybrid geodesic optimization framework for this optimization problem, and demonstrate that Riemannian gradient descent converges at a rate of $\mathcal{O}(1/T)$ under the hybrid framework.

A potential future direction could be generalizing this result for computing quantum Rényi divergences.
Our framework can also be applied to other problems.
For instance, the objective functions in the Kelly criterion in mathematical finance \cite{kelly1956new} and in computing the John Ellipsoid \cite{cohen2019near} are both g-smooth with respect to the Poincaré metric.
Another interesting direction is to develop accelerated or stochastic gradient descent-like algorithms in this hybrid scenario.
\section*{Acknowledgments}

{\small G.-R.~Wang, C.-E.~Tsai, and Y.-H.~Li are supported by the Young Scholar Fellowship (Einstein Program) of the National Science and Technology Council of Taiwan under grant number NSTC 112-2636-E-002-003, by the 2030 Cross-Generation Young Scholars Program (Excellent Young Scholars) of the National Science and Technology Council of Taiwan under grant number NSTC 112-2628-E-002-019-MY3, by the research project “Pioneering Research in Forefront Quantum Computing, Learning and Engineering” of National Taiwan University under grant numbers NTU-CC-112L893406 and NTU-CC-113L891606, and by the Academic Research-Career Development Project (Laurel Research Project) of National Taiwan University under grant numbers NTU-CDP-112L7786 and NTU-CDP-113L7763.

H.-C.~Cheng~is supported by the Young Scholar Fellowship (Einstein Program) of the National Science and Technology Council, Taiwan (R.O.C.) under Grants No.~NSTC 112-2636-E-002-009, No.~NSTC 112-2119-M-007-006, No.~NSTC 112-2119-M-001-006, No.~NSTC 112-2124-M-002-003, by the Yushan Young Scholar Program of the Ministry of Education (MOE), Taiwan (R.O.C.) under Grants No.~NTU-112V1904-4, and by the research project ``Pioneering Research in Forefront Quantum Computing, Learning and Engineering'' of National Taiwan University under Grant No.~NTC-CC-113L891605.
H.-C.~Cheng acknowledges support from the ``Center for Advanced Computing and Imaging in Biomedicine (NTU-113L900702)'' by the MOE in Taiwan.}

\bibliographystyle{plainnat}
\bibliography{bibliography}

\begin{thebibliography}{50}
\providecommand{\natexlab}[1]{#1}
\providecommand{\url}[1]{\texttt{#1}}
\expandafter\ifx\csname urlstyle\endcsname\relax
  \providecommand{\doi}[1]{doi: #1}\else
  \providecommand{\doi}{doi: \begingroup \urlstyle{rm}\Url}\fi

\bibitem[Absil et~al.(2008)Absil, Mahony, and Sepulchre]{absil2008optimization}
P-A Absil, Robert Mahony, and Rodolphe Sepulchre.
\newblock \emph{Optimization Algorithms on Matrix Manifolds}.
\newblock Princeton Univ. Press, 2008.

\bibitem[Antonakopoulos et~al.(2020)Antonakopoulos, Belmega, and
  Mertikopoulos]{antonakopoulos2020online}
Kimon Antonakopoulos, Elena~Veronica Belmega, and Panayotis Mertikopoulos.
\newblock Online and stochastic optimization beyond {L}ipschitz continuity: A
  {Riemannian} approach.
\newblock In \emph{Int. Conf. Learning Representations}, 2020.

\bibitem[Arimoto(1973)]{arimoto73}
S.~Arimoto.
\newblock On the converse to the coding theorem for discrete memoryless
  channels (corresp.).
\newblock \emph{IEEE Trans. Inf. Theory}, 19\penalty0 (3):\penalty0 357--359,
  May 1973.
\newblock ISSN 0018-9448.
\newblock \doi{10.1109/TIT.1973.1055007}.

\bibitem[Augustin(1969)]{augustin69}
U.~Augustin.
\newblock Error estimates for low rate codes.
\newblock \emph{Zeitschrift f{\"u}r {W}ahrscheinlichkeitstheorie und
  {V}erwandte {G}ebiete}, 14\penalty0 (1):\penalty0 61--88, March 1969.
\newblock ISSN 0178-8051.
\newblock \doi{10.1007/BF00534118}.

\bibitem[Augustin(1978)]{augustin1978noisy}
Udo Augustin.
\newblock Noisy channels.
\newblock \emph{Habilitation Thesis, Universität Erlangen-Nürnberg}, 1978.

\bibitem[Bhatia et~al.(2019)Bhatia, Jain, and Lim]{bhatia2019bures}
Rajendra Bhatia, Tanvi Jain, and Yongdo Lim.
\newblock On the {Bures--Wasserstein} distance between positive definite
  matrices.
\newblock \emph{Expo. Math.}, 37\penalty0 (2):\penalty0 165--191, 2019.

\bibitem[Blahut(1974)]{Bla74}
Richard Blahut.
\newblock Hypothesis testing and information theory.
\newblock \emph{IEEE Trans. Inf. Theory}, 20\penalty0 (4):\penalty0 405--417,
  July 1974.

\bibitem[Boumal(2023)]{boumal2023introduction}
Nicolas Boumal.
\newblock \emph{An Introduction to Optimization on Smooth Manifolds}.
\newblock Cambridge Univ. Press, 2023.

\bibitem[Bubeck(2015)]{bubeck2015convex}
S{\'e}bastien Bubeck.
\newblock Convex optimization: {Algorithms} and complexity.
\newblock \emph{Found. Trends Mach. Learn.}, 8\penalty0 (3-4):\penalty0
  231--357, November 2015.

\bibitem[Chen et~al.(2017)Chen, He, Jagmohan, and
  Lastras-Monta{\~n}o]{chen_reliability_2017}
Jun Chen, Da-ke He, Ashish Jagmohan, and Luis~A Lastras-Monta{\~n}o.
\newblock On the reliability function of variable-rate {Slepian-Wolf} coding.
\newblock \emph{Entropy}, 19\penalty0 (8):\penalty0 389, July 2017.

\bibitem[Cheng and Nakibo{\u{g}}lu(2021)]{CN21}
Hao-Chung Cheng and Bar{\i}{\c{s}} Nakibo{\u{g}}lu.
\newblock On the existence of the {Augustin} mean.
\newblock In \emph{2021 IEEE Information Theory Workshop (ITW)}, pages 1--6,
  2021.

\bibitem[Cheng et~al.(2019)Cheng, Hsieh, and Tomamichel]{CHT19}
Hao-Chung Cheng, Min-Hsiu Hsieh, and Marco Tomamichel.
\newblock Quantum sphere-packing bounds with polynomial prefactors.
\newblock \emph{{IEEE} Trans. Inf. Theory}, 65\penalty0 (5):\penalty0
  2872--2898, January 2019.
\newblock \doi{10.1109/tit.2019.2891347}.

\bibitem[Cheng et~al.(2022{\natexlab{a}})Cheng, Gao, and Hsieh]{CGH18}
Hao-Chung Cheng, Li~Gao, and Min-Hsiu Hsieh.
\newblock Properties of noncommutative {R}{\'{e}}nyi and {Augustin}
  information.
\newblock \emph{Commun. Math. Phys.}, 390\penalty0 (2):\penalty0 501--544,
  March 2022{\natexlab{a}}.

\bibitem[Cheng et~al.(2022{\natexlab{b}})Cheng, Hanson, Datta, and
  Hsieh]{CHDH22}
Hao-Chung Cheng, Eric~P Hanson, Nilanjana Datta, and Min-Hsiu Hsieh.
\newblock Duality between source coding with quantum side information and
  classical-quantum channel coding.
\newblock \emph{IEEE Trans. Inf. Theory}, 68\penalty0 (11):\penalty0
  7315--7345, June 2022{\natexlab{b}}.

\bibitem[Chewi et~al.(2020)Chewi, Maunu, Rigollet, and
  Stromme]{chewi2020gradient}
Sinho Chewi, Tyler Maunu, Philippe Rigollet, and Austin~J Stromme.
\newblock Gradient descent algorithms for {Bures-Wasserstein} barycenters.
\newblock In \emph{Conf. Learning Theory}, pages 1276--1304, 2020.

\bibitem[Cohen et~al.(2019)Cohen, Cousins, Lee, and Yang]{cohen2019near}
Michael~B Cohen, Ben Cousins, Yin~Tat Lee, and Xin Yang.
\newblock A near-optimal algorithm for approximating the {John} {Ellipsoid}.
\newblock In \emph{Conf. Learning Theory}, pages 849--873, 2019.

\bibitem[{Csisz\'ar}(1995)]{Csi95}
Imre {Csisz\'ar}.
\newblock Generalized cutoff rates and {R\'enyi's} information measures.
\newblock \emph{IEEE Trans. Inf. Theory}, 41\penalty0 (1):\penalty0 26--34,
  January 1995.
\newblock \doi{10.1109/18.370121}.

\bibitem[Csisz{\'a}r and K{\"o}rner(2011)]{CK11}
Imre Csisz{\'a}r and J{\'a}nos K{\"o}rner.
\newblock \emph{Information Theory: Coding Theorems for Discrete Memoryless
  Systems}.
\newblock Cambridge Univ. Press, 2011.

\bibitem[Dalai and Winter(2017)]{DW14}
Marco Dalai and Andreas Winter.
\newblock Constant compositions in the sphere packing bound for
  classical-quantum channels.
\newblock \emph{IEEE Trans. Inf. Theory}, 63\penalty0 (9):\penalty0 5603--5617,
  July 2017.

\bibitem[Dueck and Korner(1979)]{dueckK79}
G.~Dueck and J.~Korner.
\newblock Reliability function of a discrete memoryless channel at rates above
  capacity (corresp.).
\newblock \emph{IEEE Trans. Inf. Theory}, 25\penalty0 (1):\penalty0 82--85,
  January 1979.
\newblock ISSN 0018-9448.
\newblock \doi{10.1109/TIT.1979.1056003}.

\bibitem[Han et~al.(2021)Han, Mishra, Jawanpuria, and Gao]{han2021riemannian}
Andi Han, Bamdev Mishra, Pratik~Kumar Jawanpuria, and Junbin Gao.
\newblock On {Riemannian} optimization over positive definite matrices with the
  {Bures-Wasserstein} geometry.
\newblock \emph{Adv. Neural Information Processing Systems}, 34:\penalty0
  8940--8953, December 2021.

\bibitem[Haroutunian(1968)]{Har68}
EA~Haroutunian.
\newblock Bounds for the exponent of the probability of error for a
  semicontinuous memoryless channel.
\newblock \emph{Problemy Peredachi Informatsii}, 4\penalty0 (4):\penalty0
  37--48, 1968.

\bibitem[Hosseini and Sra(2015)]{hosseini2015matrix}
Reshad Hosseini and Suvrit Sra.
\newblock Matrix manifold optimization for {Gaussian} mixtures.
\newblock \emph{Adv. Neural Information Processing Systems}, 28, 2015.

\bibitem[Hosseini and Sra(2020)]{hosseini2020alternative}
Reshad Hosseini and Suvrit Sra.
\newblock An alternative to {EM} for {Gaussian} mixture models: batch and
  stochastic {Riemannian} optimization.
\newblock \emph{Math. Program.}, 181\penalty0 (1):\penalty0 187--223, May 2020.

\bibitem[Kelly(1956)]{kelly1956new}
John~L Kelly.
\newblock A new interpretation of information rate.
\newblock \emph{Bell Syst. Tech. J.}, 35\penalty0 (4):\penalty0 917--926, July
  1956.

\bibitem[Lan(2020)]{lan2020first}
Guanghui Lan.
\newblock \emph{{First-order and Stochastic Optimization Methods for Machine
  Learning}}.
\newblock Springer, 2020.

\bibitem[Lee(2018)]{lee2018introduction}
John~M Lee.
\newblock \emph{Introduction to {Riemannian} Manifolds}.
\newblock Springer, 2018.

\bibitem[Li and Cevher(2019)]{li2019convergence}
Yen-Huan Li and Volkan Cevher.
\newblock Convergence of the exponentiated gradient method with {Armijo} line
  search.
\newblock \emph{J. Optim. Theory Appl.}, 181\penalty0 (2):\penalty0 588--607,
  May 2019.

\bibitem[Mojahedian et~al.(2019)Mojahedian, Beigi, Gohari, Yassaee, and
  Aref]{mojahedian2019correlation}
Mohammad~Mahdi Mojahedian, Salman Beigi, Amin Gohari, Mohammad~Hossein Yassaee,
  and Mohammad~Reza Aref.
\newblock A correlation measure based on vector-valued {$L_p$}-norms.
\newblock \emph{IEEE Trans. Inf. Theory}, 65\penalty0 (12):\penalty0
  7985--8004, August 2019.

\bibitem[Mosonyi and Ogawa(2017)]{mosonyiO17}
M.~Mosonyi and T.~Ogawa.
\newblock Strong converse exponent for classical-quantum channel coding.
\newblock \emph{Commun. Math. Phys.}, 355\penalty0 (1):\penalty0 373--426,
  October 2017.
\newblock ISSN 1432-0916.
\newblock \doi{10.1007/s00220-017-2928-4}.

\bibitem[Mosonyi and Ogawa(2021)]{mosonyiO21}
M.~Mosonyi and T.~Ogawa.
\newblock Divergence radii and the strong converse exponent of
  classical-quantum channel coding with constant compositions.
\newblock \emph{IEEE Trans. Inf. Theory}, 67\penalty0 (3):\penalty0 1668--1698,
  December 2021.

\bibitem[Nakibo{\u{g}}lu(2019)]{nakibouglu2019augustin}
Bar{\i}{\c{s}} Nakibo{\u{g}}lu.
\newblock The {Augustin} capacity and center.
\newblock \emph{Probl. Inf. Transm.}, 55:\penalty0 299--342, October 2019.

\bibitem[Nakibo{\u{g}}lu(2020)]{nakibouglu2020simple}
BARI{\c{S}} Nakibo{\u{g}}lu.
\newblock A simple derivation of the refined sphere packing bound under certain
  symmetry hypotheses.
\newblock \emph{Turk. J. Math.}, 44\penalty0 (3):\penalty0 919--948, 2020.

\bibitem[Nesterov(2018)]{nesterov2018lectures}
Yurii Nesterov.
\newblock \emph{Lectures on Convex Optimization}.
\newblock Springer, 2018.

\bibitem[Omura(1975)]{omura75}
J.~K. Omura.
\newblock A lower bounding method for channel and source coding probabilities.
\newblock \emph{Inf. Control.}, 27\penalty0 (2):\penalty0 148--177, February
  1975.
\newblock ISSN 0019-9958.
\newblock \doi{10.1016/S0019-9958(75)90120-5}.

\bibitem[Pennec(2020)]{pennec2020manifold}
Xavier Pennec.
\newblock Manifold-valued image processing with {SPD} matrices.
\newblock In \emph{Riemannian Geometric Statistics in Medical Image Analysis},
  pages 75--134. Elsevier, 2020.

\bibitem[Pennec et~al.(2006)Pennec, Fillard, and Ayache]{pennec2006riemannian}
Xavier Pennec, Pierre Fillard, and Nicholas Ayache.
\newblock A {Riemannian} framework for tensor computing.
\newblock \emph{Int. J. Comput. Vis.}, 66:\penalty0 41--66, January 2006.

\bibitem[Rapcsak(1991)]{rapcsak1991geodesic}
Tamas Rapcsak.
\newblock Geodesic convexity in nonlinear optimization.
\newblock \emph{J. Optim. Theory Appl.}, 69\penalty0 (1):\penalty0 169--183,
  April 1991.

\bibitem[R{\'e}nyi(1961)]{Ren62}
Alfr{\'e}d R{\'e}nyi.
\newblock On measures of entropy and information.
\newblock In \emph{Proc. 4th Berkeley Symp. Mathematical Statistics and
  Probability, Volume 1: Contributions to the Theory of Statistics}, volume~4,
  pages 547--562, 1961.

\bibitem[Shen et~al.(2023)Shen, Gao, and Cheng]{SGC23}
Yu-Chen Shen, Li~Gao, and Hao-Chung Cheng.
\newblock Privacy amplification against quantum side information via regular
  random binning.
\newblock In \emph{2023 59th Annu. Allerton Conference on Communication,
  Control, and Computing (Allerton)}, pages 1--8, 2023.

\bibitem[Udriste(1994)]{udriste1994convex}
Constantin Udriste.
\newblock \emph{Convex functions and optimization methods on {Riemannian}
  manifolds}.
\newblock Springer Science \& Business Media, 1994.

\bibitem[van Erven and Harremoes(2014)]{EH14}
Tim van Erven and Peter Harremoes.
\newblock R{\'{e}}nyi divergence and {Kullback-Leibler} divergence.
\newblock \emph{IEEE Trans. Inf. Theory}, 60\penalty0 (7):\penalty0 3797--3820,
  June 2014.
\newblock \doi{10.1109/tit.2014.2320500}.

\bibitem[Verdu(2015)]{Ver15}
Sergio Verdu.
\newblock {$\alpha$}-mutual information.
\newblock In \emph{2015 Information Theory and Applications Workshop (ITA)},
  pages 1--6, 2015.

\bibitem[Verd{\'u}(2021)]{Ver21}
Sergio Verd{\'u}.
\newblock Error exponents and $\alpha$-mutual information.
\newblock \emph{Entropy}, 23\penalty0 (2):\penalty0 199, August 2021.

\bibitem[Vishnoi(2018)]{vishnoi2018geodesic}
Nisheeth~K Vishnoi.
\newblock Geodesic convex optimization: {Differentiation} on manifolds,
  geodesics, and convexity.
\newblock 2018.
\newblock arXiv:1806.06373.

\bibitem[Weber and Sra(2023)]{weber2023global}
Melanie Weber and Suvrit Sra.
\newblock Global optimality for {Euclidean} {CCCP} under {Riemannian}
  convexity.
\newblock In \emph{Int. Conf. Machine Learning}, pages 36790--36803, 2023.

\bibitem[Wiesel(2012)]{wiesel2012geodesic}
Ami Wiesel.
\newblock Geodesic convexity and covariance estimation.
\newblock \emph{IEEE Trans. Signal Process.}, 60\penalty0 (12):\penalty0
  6182--6189, September 2012.

\bibitem[You et~al.(2022)You, Cheng, and Li]{you2022minimizing}
Jun-Kai You, Hao-Chung Cheng, and Yen-Huan Li.
\newblock Minimizing quantum {R{\'e}nyi} divergences via mirror descent with
  {Polyak} step size.
\newblock In \emph{2022 IEEE Int. Symp. Information Theory (ISIT)}, pages
  252--257, 2022.

\bibitem[Zhang and Sra(2016)]{zhang2016first}
Hongyi Zhang and Suvrit Sra.
\newblock First-order methods for geodesically convex optimization.
\newblock In \emph{Conf. Learning Theory}, pages 1617--1638, 2016.

\bibitem[Zhang et~al.(2013)Zhang, Wiesel, and Greco]{zhang2013multivariate}
Teng Zhang, Ami Wiesel, and Maria~Sabrina Greco.
\newblock Multivariate generalized {Gaussian} distribution: {Convexity} and
  graphical models.
\newblock \emph{IEEE Trans. Signal Process.}, 61\penalty0 (16):\penalty0
  4141--4148, June 2013.

\end{thebibliography}

\appendix
\addcontentsline{toc}{section}{Appendices}
\section*{Appendices}

\section{Proof of Lemma \ref{lemma:asymptotic convergence}}\label{appendix:asymptotic convergence proof}

Since $x_\infty\in\mathcal{M}$, by Lemma \ref{lemma:vanishing grad}, we have
\begin{equation*}
    \Vert \grad_{\mathfrak{h}} f(x_\infty)\Vert_{\mathfrak{h}(x_\infty)}=0.
\end{equation*}
Note that $\grad_{\mathfrak{h}}f(x)$ is the (Riesz) representation of the differential form $\mathrm{d}f(x)$ on the tangent space $T_x\mathcal{M}$ with respect to the metric $\mathfrak{h}$, and $\grad_{\mathfrak{g}}f(x)$ is the representation of the \emph{same} differential form $\mathrm{d}f(x)$ under another metric $\mathfrak{g}$.
Therefore, for every $v\in T_x\mathcal{M}$, we have
\begin{equation*}
    \langle\grad_{\mathfrak{h}}f(x), v\rangle_{\mathfrak{h}(x)}    =\mathrm{d}f(x)[v]    =\langle\grad_{\mathfrak{g}}f(x), v\rangle_{\mathfrak{g}(x)},
\end{equation*}
which implies
\begin{equation*}
    \Vert \grad_{\mathfrak{g}} f(x_\infty)\Vert_{\mathfrak{g}(x_\infty)}=0.
\end{equation*}
Since $f$ is g-convex with respect to $\mathfrak{g}$, having zero Riemannian gradient with respect to $\mathfrak{g}$ at $x$ is equivalent to $x$ being a global minimizer of $f$ on $\mathcal{M}$.
\section{Proof of Theorem \ref{thm:main}}\label{appendix:main_proof}

We have shown that
\begin{equation*}
    \delta_{t+1}-\delta_t\leq-\frac{1}{2L}\frac{\delta_t^2}{\Vert \log_{\mathfrak{g}(x_t)}x_\infty\Vert^2_{\mathfrak{h}(x_t)}}.
\end{equation*}

Let $w_t:=\frac{1}{\Vert \log_{\mathfrak{g}(x_t)}x_\infty\Vert^2_{\mathfrak{h}(x_t)}}$ and divide both sides by $\delta_t\delta_{t+1}$.
Since the sequence $\{ \delta_t \}$ is non-increasing, we write
\begin{equation*}
    \frac{1}{\delta_t}-\frac{1}{\delta_{t+1}}\leq-\frac{w_t}{2L}\frac{\delta_t}{\delta_{t+1}}\leq-\frac{w_t}{2L}.
\end{equation*}
By a telescopic sum, we get
\begin{equation*}
    -\frac{1}{\delta_{t+1}}\leq\frac{1}{\delta_1}-\frac{1}{\delta_{t+1}}\leq-\frac{\sum_t w_t}{2L}.
\end{equation*}
Therefore, we have
\begin{equation*}
    f(x_{T+1})-f(x_\infty)\leq\frac{2L}{\sum w_t}\leq\frac{2L}{T}\sup_{t\in\mathbb{N}}\Vert \log_{\mathfrak{g}(x_t)}x_\infty\Vert^2_{\mathfrak{h}(x_t)}.
\end{equation*}
\section{Proof of Lemma \ref{lemma:relaxation}}\label{appendix:relaxation proof}

Given any vector $x\in\RDp$, we can write it as $x = \lambda y$ where $\lambda=\sum_{i\in[N]}x^{(i)}$ and $y\in\simplex$.
We have
\begin{equation*}
    g_\alpha(x)=g_\alpha(\lambda y)=\lambda-\log\lambda+ f_\alpha(y).
\end{equation*}
Observe that 
\begin{equation*}
    \frac{\partial}{\partial \lambda}g_\alpha=1-\frac{1}{\lambda}
\end{equation*}
and $g_a(\lambda y)$ is convex in $\lambda$.
This means for every $y\in\simplex$, the function $g_\alpha(\lambda y)$ achieves its minimum at $\lambda=1$.
Therefore, let $y^\star$ be a minimizer of $f_\alpha$ on $\simplex$, for every $\lambda>0$ and $y\in\simplex$, we have
\begin{align*}
    f_\alpha(y^\star) &\leq f_\alpha(y)\\
    \Leftrightarrow 1+f_\alpha(y^\star) &\leq 1+f_\alpha(y) \\
    \Leftrightarrow g_\alpha(y^\star) &\leq g_\alpha(y) \leq g_\alpha(\lambda y).
\end{align*}
This shows that $y^\star$ is a minimizer of $f_\alpha(y)$ on $\simplex$ if and only if $x^\star=y^\star$ is a minimizer of $g_\alpha(x)$.
\section{Proof of Lemma \ref{lemma:AI g-smoothness}}\label{appendix:AI g-smoothness proof}

We will utilize the following equivalent definition of g-convexity and g-smoothness.

\begin{Def}[\cite{zhang2016first, han2021riemannian}]
    For a function $f:\mathcal{M}\to\mathbb{R}$ defined on $(\mathcal{M}, \mathfrak{g})$, we say that $f$ is g-convex on ($\mathcal{M}, \mathfrak{g}$) if for every geodesic $\gamma(t):[0, 1]\to\mathcal{M}$ connecting any two points $x, y\in\mathcal{M}$, $f(\gamma(t))$ is convex in $t$ in the Euclidean sense.
    We say that $f$ is g-$L$-smooth ($\mathcal{M}, \mathfrak{g}$) if for every geodesic $\gamma(t):[0, 1]\to\mathcal{M}$ connecting any two points $x, y\in\mathcal{M}$, $f(\gamma(t))$ is $Ld^2(x, y)$-smooth in $t$ in the Euclidean sense.
\end{Def}\vspace{1ex}

Let $\gamma(t)$ be the geodesic connecting $x$ and $y$ for $x, y\in\RDpp$.
If $f_p(\gamma(t))$ is $Ld(x, y)$-smooth in the Euclidean sense, then the function $f_\alpha(\gamma(t))=\mathbb{E}_p f_p(\gamma(t))$ is also $Ld(x, y)$-smooth in the Euclidean sense.
Therefore, it suffices to prove the g-smoothness for the function $f_p(x):=-\frac{1}{1-\alpha}\log\langle p^\alpha, x^{1-\alpha}\rangle$ for any $p\in\RDp$.

We now prove that
\begin{equation*}
    f_p(\gamma(t))=-\frac{1}{1-\alpha}\log\langle p^\alpha, \gamma(t)^{1-\alpha}\rangle
\end{equation*}
is $|1-\alpha|d^2(x, y)$-smooth in the Euclidean sense.
Note that 
\begin{equation*}
    \gamma'(t)=\gamma(t)\odot \log(y\oslash x).
\end{equation*}
We have
\begin{equation*}
    \frac{\du}{\du t}f_p(\gamma(t))=-\frac{\langle p^\alpha, \gamma(t)^{1-\alpha}\odot\log(y\oslash x)\rangle}{\langle p^\alpha, \gamma(t)^{1-\alpha}\rangle}.
\end{equation*}
The second-order derivative is
\begin{align*}
    \frac{\du^2}{{\du t}^2}f_p(\gamma(t))
    &=(1-\alpha)\bigg[\underbrace{\left(\frac{\langle p^\alpha\odot\gamma(t)^{1-\alpha}, \log(y\oslash x)\rangle}{\langle p^\alpha, \gamma(t)^{1-\alpha}\rangle}\right)^2}_{\text{(1)}}\\
    &- \underbrace{\frac{\langle p^\alpha\odot\gamma(t)^{1-\alpha}, (\log(y\oslash x))^2\rangle}{\langle p^\alpha, \gamma(t)^{1-\alpha}\rangle}}_{\text{(2)}}\bigg].
\end{align*}
Note that $p, (\log(y\oslash x))^2\in\RDp$ and $\gamma(t)\in\RDpp$, so the quantities (1) and (2) are both non-negative.
Our strategy is to show that both quantities (1) and (2) are upper bounded by $d^2(x, y)$, which then implies
\begin{equation*}
    \frac{\du^2}{{\du t}^2}f_p(\gamma(t))\in [-|1-\alpha|d^2(x, y), |1-\alpha|d^2(x, y)].
\end{equation*}
This gives the desired result by the fact that a function is $L$-smooth if and only if its Hessian is upper bounded by $LI$, where $I$ denotes the identity matrix.

\subsection{Upper Bound of the quantity (1)}
The numerator of the quantity (1) is 
\begin{align*}
    \langle p^\alpha\odot\gamma(t)^{1-\alpha}&, \log(y\oslash x)\rangle^2\\
    &\leq \left\Vert p^\alpha\odot\gamma(t)^{1-\alpha} \right\Vert^2_2 \cdot \left\Vert \log(y\oslash x))\right\Vert^2_2\\
    &\leq\langle p^\alpha, \gamma(t)^{1-\alpha} \rangle^2 \left\Vert \log(y\oslash x))\right\Vert^2_2\\
    &=\langle p^\alpha, \gamma(t)^{1-\alpha} \rangle^2 d^2(x, y).
\end{align*}
The first inequality is due to the Cauchy-Schwarz inequality, and the second inequality holds because $\sum (x^{(i)})^2\leq (\sum x^{(i)})^2$ when $x^{(i)}\geq 0$ for all $i\in[N]$.
Thus, the quantity (1) is upper bounded by $d^2(x, y)$.

\subsection{Upper Bound of the quantity (2)}
The numerator of the quantity (2) is
\begin{align*}
    \langle p^\alpha\odot \gamma(t)^{1-\alpha}&, (\log(y\oslash x))^2\rangle \\
    &\leq \left\Vert p^\alpha\odot\gamma(t)^{1-\alpha} \right\Vert_2 \cdot \left\Vert  (\log(y\oslash x))^2\right\Vert_2 \\
    &\leq\langle p^\alpha, \gamma(t)^{1-\alpha}\rangle\left\Vert \log(y\oslash x))\right\Vert_2^2\\
    &=\langle p^\alpha, \gamma(t)^{1-\alpha}\rangle d^2(x, y).
\end{align*}
The first inequality is due to the Cauchy-Schwarz inequality, and the second inequality is because $\sum (x^{(i)})^2\leq (\sum x^{(i)})^2$ when $x^{(i)}\geq 0$ for all $i\in[N]$.
Thus, the quantity (2) is upper bounded by $d^2(x, y)$.

\subsection{Geodesic Smoothness of $g_\alpha$}
We only need to show that $h(x): x \mapsto \langle \mathbbm{1}, x\rangle$ is g-1-smooth.
By direct calculation, we have
\begin{align*}
    \frac{\du^2}{{\du t}^2}h(\gamma(t))&=\sum (\gamma(t))^{(i)}\left(\log\frac{x^{(i)}}{y^{(i)}}\right)^2\\
    &\leq\sum\left(\log\frac{x^{(i)}}{y^{(i)}}\right)^2=d^2(x, y),
\end{align*}
where the inequality is due to the concavity and monotonicity of $\log$, which implies
\begin{equation*}
    (\gamma(t))^{(i)}=(x^{(i)})^{1-t}(y^{(i)})^t\leq (1-t)x^{(i)}+ty^{(i)}\leq 1.
\end{equation*}

\section{Proof of Lemma \ref{lemma:remain in box}}\label{appendix:remain in box proof}

Let $x\leq \mathbbm{1}$, then the next iterate is 
\begin{equation*}
        x_+=x\odot\exp\left(-\frac{1}{|1-\alpha|+1}x\odot\nabla g_\alpha(x)\right).
\end{equation*}
A direct calculation shows that $x\odot-\nabla f_\alpha(x)\in\simplex$, so
\begin{equation*}
    -x^{(i)}\odot\nabla^{(i)} f_\alpha(x)\leq 1,
\end{equation*}
where $\nabla^{(i)} f(x)$ is the $i$-th component of $\nabla f(x)$.
By the inequality $\log(a)\leq b(a^{1/b}-1)$ for any $a, b>0$, we have
\begin{align*}
    \log x^{(i)}&\leq\frac{1}{|1-\alpha|+1}\left((x^{(i)})^{|1-\alpha|+1}-1\right)\\
    &\leq\frac{1}{|1-\alpha|+1}(x^{(i)}-1)\\
    &\leq\frac{1}{|1-\alpha|+1}\left(x^{(i)}+x^{(i)}\nabla^{(i)} f_\alpha(x)\right)\\
    &=\frac{1}{|1-\alpha|+1}\left(x^{(i)}\nabla^{(i)} g_\alpha(x)\right).
\end{align*}
Therefore,
\begin{align*}
    x^{(i)}\leq \exp\left(\frac{1}{|1-\alpha|+1} \left(x^{(i)}\nabla^{(i)} g_\alpha(x)\right)\right).
\end{align*}
Hence, 
\begin{equation*}
    x^{(i)}_+=x^{(i)} \exp\left(-\frac{1}{|1-\alpha|+1} \left(x^{(i)}\nabla^{(i)} g_\alpha(x)\right)\right)\leq 1.
\end{equation*}
This gives the desired result.
\section{Proof of Lemma \ref{lemma:boundary issue}}\label{appendix:boundary issue proof}

If the limit $x_\infty$ lies in $\RDpp$, then it is a minimizer of $f(x)$ by Theorem \ref{thm:main}.
Assume that the limit point $x_\infty\in\partial\RDpp$ and $x_\infty\notin\argmin_{x\in\RDp}f(x)$.
Then, by the first-order optimality condition, there exists some $x\in\RDp$ such that
\begin{equation}\label{eq:foc}
    \langle \nabla f(x_\infty), x-x_\infty\rangle=\sum_{i\in[N]}\nabla^{(i)} f(x_\infty)(x^{(i)}-x^{(i)}_\infty)<0,
\end{equation}
where $\nabla^{(i)} f(x)$ is the $i$-th component of the vector $\nabla f(x)$.

By Lemma \ref{lemma:vanishing grad}, we have
\begin{equation*}
    \Vert \grad f(x_\infty)\Vert_{x_\infty}=\Vert x_\infty\odot\nabla f(x_\infty)\Vert_2=0,
\end{equation*}
where the Riemannian gradient is with respect to the Poincaré metric.
Therefore,
\begin{equation*}
    x^{(i)}_\infty\neq 0\Rightarrow \nabla^{(i)} f(x_\infty)=0.
\end{equation*}
Combining this and the inequality \eqref{eq:foc}, we get
\begin{equation*}
    \sum_{i\in[N], x^{(i)}_\infty=0} \nabla^{(i)} f(x_\infty)x^{(i)} < 0,
\end{equation*}
which means there exists some $j\in[N]$ such that
\begin{equation*}
    \nabla^{(j)} f(x_\infty)<0\text{ and }x^{(j)}_\infty=0.
\end{equation*}
By the continuity of $\nabla f(x)$, there exists an open neighborhood $U\subset\RD$ of $x_\infty$ such that for all $x\in U$
\begin{equation*}
    \nabla^{(j)} f(x)<0.
\end{equation*}
Since $x_\infty$ is the limit point of the sequence $\{x_t\}$, there exists $T>0$ such that $x_t\in U$ for all $t>T$, that is,
\begin{equation*}
    \nabla^{(j)} f(x_t)<0.
\end{equation*}
However, this implies
\begin{equation*}
    -\frac{1}{L}x_t^{(j)}\nabla^{(j)} f(x_t)>0\quad \forall t>T,
\end{equation*}
since $x_t\in\RDpp$ and $\nabla^{(j)} f(x_t)<0$.
Therefore, by the iteration rule of RGD, we have
\begin{equation*}
    x_{t+1}^{(j)}=x_{t}^{(j)} \exp\left(-\frac{1}{L}x_{t}^{(j)}\nabla^{(j)} f(x_t)\right)>x_{t}^{(j)}.
\end{equation*}
This shows that $\{x_t^{(j)}\}$ is increasing after $t>T$, which means this sequence cannot converge to $0$.
This contradicts our assumption.
\section{Proof of Lemma \ref{prop:applied to f}}\label{appendix:applied to f proof}
Since $g_\alpha$ is convex and g-smooth with respect to the Poincaré metric, by Theorem \ref{thm:main} and Lemma \ref{lemma:boundary issue}, we have
\begin{equation*}
    g_\alpha(x_{T+1})-g_\alpha(x^\star)\leq\frac{2(|1-\alpha|+1)}{T}\sup_{t\in\mathbb{N}}\Vert x^\star\oslash x_t-\mathbbm{1}\Vert_2^2.
\end{equation*}
Note that for any $\lambda >0$ and a fixed $x\in\simplex$, the function $g_\alpha(\lambda x)$ is minimized when $\lambda=1$, as shown in Appendix \ref{appendix:relaxation proof}.
Therefore, $x^\star\in\simplex$, and we have
\begin{align*}
    f_\alpha(\overline{x}_{T+1})-f_\alpha(x^\star)
    &=g_\alpha(\overline{x}_{T+1})-g_\alpha(x^\star)\\
    &\leq g_\alpha(x_{T+1})-g_\alpha(x^\star).
\end{align*}

\end{document}